\documentclass{article}
\usepackage{bm}
\usepackage{graphicx}

\begin{document}
\title{Adaptability and Diversity in Simulated Turn-taking Behaviour}
\author{Hiroyuki Iizuka \and Takashi Ikegami}
\date{}
\maketitle
\vspace{-1cm}
\begin{center}
Department of General Systems Sciences, \\
The Graduate School of Arts and Sciences, University of Tokyo,\\
3-8-1 Komaba, Tokyo 153-8902, Japan\\
\end{center}

\begin{abstract}

Turn-taking behaviour is simulated in
a coupled agents system. Each agent is modelled as a mobile robot with two wheels.
A recurrent neural network is used to produce the motor outputs and to
hold the internal dynamics.
Agents are developed to take turns on a
two-dimensional arena by causing the network structures to evolve.

Turn-taking is established using either regular or chaotic behaviour of the agents.
It is found that
chaotic turn-takers are more sensitive to the adaptive inputs from the other agent.
Conversely, regular turn-takers are comparatively robust against noisy inputs,
owing to their restricted dynamics. From many observations, including
turn-taking with virtual agents, we claim that
there is a complementary relationship between robustness and adaptability.
Furthermore, by investigating the recoupling of agents from different GA
generations, we report the emergence of a new turn-taking behaviour.
Potential for synthesizing a new form of motion is another characteristic of chaotic turn-takers.

\end{abstract}

{\bf Keywords:} turn-taking, adaptive behavior, diversity of behaviors,
cognition, embodiment

\section{Introduction}

Dynamical systems can theoretically simulate
behaviour produced over time with interactions between various
entities.
This approach, based on embodied cognition \cite{acper,evorobo,understanding}, has a different
perspective from the traditional AI approaches. That is, representations
are not given as symbols in advance but are only realized, by the
dynamics, over time \cite{beer1997,pollack,tani}. Cognitive structure is characterized by
geometrical and flow patterns in an adequate phase
space. As well as being characterized by attractor types (e.g., fixed point, limit-cycle,
and strange attractors) they are also characterized by chaotic itinerancy and other novel concepts,
such as open-ended evolution/dynamics, that describe their inherent behaviour.

Richness and the potential of the dynamical systems approach encourage us to
go beyond merely adaptive behaviour.
The higher functions, such as intention, motivation, emotion and
consciousness, are within the scope of this study.
Grey Walter has started the discussion of emotional, or play-like, behaviour
by synthesizing artificial creatures \cite{walter,walter51}. A wheeled vehicle containing a simple electric
circuit can show unexpected and complex behaviour, comparable to that of living creatures.
Without making real robots, Braitenberg made conceptual robots to discuss the
higher functions \cite{vehicles}.
In his thought experiments, he designed vehicles using simple hard-wired
electrical connections from sensory inputs to motor outputs.
His vehicles gradually showed more complex cognitive
behaviours by providing more complex internal structures. For
example, the most primitive behaviour is a sense of ``aggression'', which is
simply given by attraction to a light source with a crossed
sensory--motor connection. However, to simulate more complex behaviour, such as
association and concept formation, he has to implement new wires, such as
mnemotorix and ergotorix wires, with some Darwinian-type selections.
Grey Walter and Braitenberg have one thing in common, in claiming that any apparently
complex
cognitive behaviour can be built up from simple sensory--motor coordination.
That is, agents can be cognitive by having physical constraints.
We basically agree that any meaningful cognition should be embodied, but focus on
different aspects.

In this paper, we focus on the cognitive behaviours of turn-taking and
imitation, caused by interactions
between two or more humans, in which it is thought that the sharing of mental states and
intentions with others is important.
There are many ways to understand psychological phenomena
by computer simulations and robot experiments rather than by studying human
behaviour directly \cite{kerstin1995,kerstin,scassellati}.
We conducted computer simulations of two
agents with internal dynamics, which were implemented by an artificial
recurrent neural network, as a model of turn-taking behaviour.
In our previous works, cognitive behaviours
were explained from the dynamical systems perspective by coupling
between agents with rich internal dynamics
\cite{timt98,timt99,ikeggenta}. 
Here, we generalize from
turn-taking behaviour to autonomous role-changing,
such as games of tag among children, and investigate the
generic underlying mechanisms using the dynamical systems method.
Therefore, this study focuses on different perspectives from those of
fixed role-playing games (e.g., a pursuit-evasion game \cite{cliff}).
Here we take turn-taking as the simplest example that shows the diversity
of dynamics.
For turn-taking behaviour, it is necessary for roles to be exchanged
autonomously, within a context constructed by the entities' behaviours, e.g.,
chaser--evader and speaker--listener.
When taking turns in a two-person conversation,
people usually avoid overlapping or interrupting each other's speech
without setting some explicit cue to switch speakers.
Some cues for this include eye contact and the detection of intonation changes.
It is considered that turn-taking is established by coordination between
predictions and the internal neural networks that compute the output from the inputs.
Therefore, coupling between agents means a coupling of anticipatory
systems with intrinsic dynamics.

By introducing neural architecture, evolutionary algorithm and a turn-taking game in \S 2 and 3,
we explore four topics in the simulation. The first topic is dynamic repertoire. We describe how
turn-taking is established with different forms of motion. In particular, we argue in \S 4.1 that regular
motion behaviour evolves into chaotic behaviour. The second topic is predictability. Each agent has to
predict the other's future behaviour one step ahead. Interestingly, prediction precision decreases
when the turn-taking role switches from one to the other. This will be discussed in \S 4.2.
The third topic is ongoingness of interactions.
Agents become robust against sensor noise; however, the turn-taking
performance is established only when agents synchronize their dynamics precisely. This point
is discussed in \S 4.3. The last topic is adaptability. As discussed in the section
on dynamic repertoire, the turn-taking pattern appears to be different for different evolutionary generations.
In section \S 4.4, we investigate the emergence of new spatio-temporal patterns by
coupling agents from different generations. In \S 5, we discuss the potential linkage between these
simulation results and the psychological experiments conducted
by C. Trevarthen \cite{tre}.
A concept of intersubjectivity is also discussed.

\section{The Model} 
We modelled the playing of a tag game in which the role of chaser, or evader, is not
given to players in advance.
There are some game models in which the roles are not predefined.
Reynolds also showed that the abilities of chasing and evading
evolve simultaneously by genetic programming in a
game of tag, which is a symmetrical pursuit-evasion game \cite{reynolds}.
The variety in the behaviour of agents adapting to their environments is
worth noting.
In Reynolds' game, switching between evader and chaser is predefined
as happening when both agents come into physical contact.
The difference between Reynolds' model and ours is the spontaneous emergence of
behaviour.
Whether an agent plays the role of a chaser or an evader is dynamically determined in our model.
On the other hand, Di Paolo modelled and studied social coordination with agents
interacting acoustically \cite{paolo}.
To avoid misperceiving the acoustical signals, their emission timings were entrained in an
anti-phase state; the resulting behaviour
resembles a turn-taking process.

There is a difference between Di Paolo's turn-taking and ours.
Both turn-taking behaviours are established
by the coordination of agents through the history of their interactions.
Di Paolo modelled turn-taking as the result of anti-phase signals to
avoid signal interference; however, we modelled turn-taking behaviour
as a result of coupling between richer internal dynamics.
Therefore, in this paper, we pay more attention to the diversity of behaviour patterns.

\subsection{Game and Environment} 
\begin{figure}[tb]
\centerline{\includegraphics[scale=0.33]{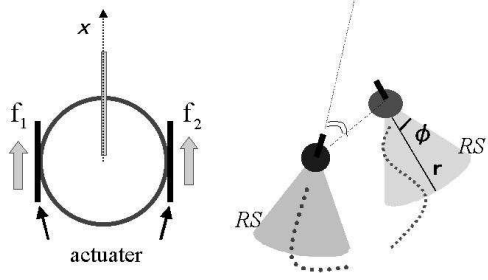}}
\caption{ Left: a schematic view of the mobile robot with
two wheels (actuators). It computes the forward force vector
and the torque strength from the force vector ($f_{1},f_{2}$) on
each actuator. Right: Two mobile robots interact to perform
turn-taking behaviour by
sensing each other's position, relative distance and heading angle.
It is robot A's turn when A enters the area that is B's rear side (RS)
position. The shape of this RS is parameterized by $r$ and $\phi$.
}
\label{fig:scope}
\end{figure}

Here each agent has a circular body of radius R, with two
diametrically opposed motors (Fig. \ref{fig:scope}).
The motors can move the agent backwards
and forwards in a two-dimensional unstructured and unlimited arena.
The motion is described by the following equation of motion in terms of an agent's
heading angle ($\theta$) and its speed ($v$) in that direction.
\begin{eqnarray}
M\dot{v}+D_{1}v+f_{1}+f_{2}=0, \\
I\ddot{\theta}+D_{2}\dot{\theta}+\tau(f_{1},f_{2})=0,
\end{eqnarray}
{\parindent 0mm
where $f_{1}$ and $f_{2}$ are the forward driving force, and $\tau$
denotes the torque.
$D_{1}$ and $D_{2}$ express the resistance coefficients, and the agents
have mass ($M$) and inertia ($I$).
We solve the equations iteratively using the Runge--Kutta method. At each time
step, the agents compute the forces from the inputs using the internal neural nets
described below.
}

We assume there is no collision between agents because we focus on the
internal states of the agents that generate turn-taking.
Two agents try to coordinate their turn-taking behaviour, each trying to get behind
the other. Because they cannot get behind each other simultaneously,
the turn-taking cannot be achieved if both agents play chaser.
Naturally, mutual turn-taking cannot be
achieved if both agents play evader either.
Therefore, it is necessary to have spontaneous symmetry break down so
that one plays the role of chaser and the other plays the role of evader.
However, mere symmetry breakdown is insufficient: temporal role changing is also required.
By using recurrent neural networks, we focus on how the turn-taking dynamics are
self-organized.

\subsection{Agent Design} 

\begin{figure}[tb]
\centerline{\includegraphics[scale=0.5]{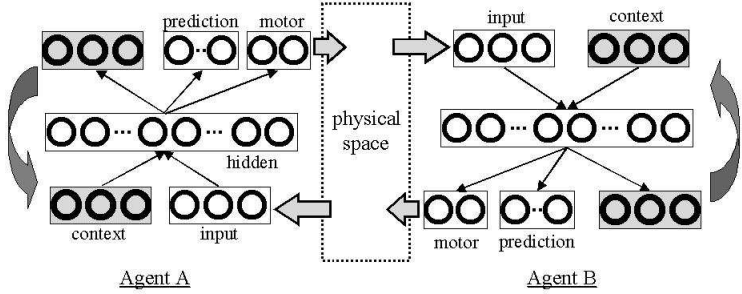}}
\caption{ Recurrent neural networks with three layers.
Input nodes receive the other agent's relative position. The final layer
consists of three types of node: context, prediction and
motor output. Context nodes feed back to the input layer. Prediction
nodes output the other's relative position in the next time step. Motor nodes
output the force vector, $f_{1}$ and $f_{2}$.}
\label{fig:rnn}
\end{figure}

We designed the agents to have recurrent neural networks (Fig. \ref{fig:rnn}).
Inputs to an agent are the other agent's position, distance and heading angle,
relative to the agent.
Agents move freely in the arena using two motors, the outputs of which are computed at
every time step of the game.
Each agent predicts the other's next
relative position, which is assigned to three output neurons.
The dynamics of the recurrent neural network are expressed by
the following equations at each time step $t$,
\begin{eqnarray}
h_{j}(t) & = & g(\sum_{i}{w_{ij}y_{i}(t)+\sum_{l}{w'_{lj}c_{l}(t-1)}}+b_{j1}),\\
z_{k}(t) & = & g(\sum_{j}{u_{jk}h_{j}(t)}+b_{j2}),\\
c_{l}(t) & = & g(\sum_{l}{u'_{jl}h_{j}(t)}+b_{j3}),\\
g(x) & = & 1/(1+\exp(-x)),
\end{eqnarray}
{\parindent 0mm
where $y_{i}, z_{k}, h_{j}$ and $c_{l}$ represent input, output, hidden and
context nodes, respectively. The respective number of nodes in these layers is set
to $(I, K, J, L) = (3, 5, 10, 3)$ throughout this paper.
The symbols $w_{ij}, u_{jk}, w'_{lj}$ and $u'_{jl}$ denote the weights from
input to hidden,
hidden to output, context to hidden, and hidden to context neurons, respectively,
and the parameter $b$ is a bias node.
In this paper, we do not consider the results of predictions,
which are discussed in \cite{ikegezca}.
This network architecture evolves using a genetic algorithm
as explained in the following section.
}

\section{Genetic Algorithm and Noisy Environment} 
\subsection{Genetic Algorithm}
We update the weights according to the turn-taking performance.
In practice, the weight set of the neural networks has a vector
representation of the real weight values, which evolve using a
genetic algorithm (GA).

We use a GA to evolve two separate populations, to avoid agents of
a single genotype from dominating, in which case
turn-taking is played between genetically similar agents.
As a result, a player has to play against itself, which we wish to avoid.
Each population contains $P$ individuals.
The performance of all $P^{2}$ paired agents from the separated populations
are evaluated at each generation.
Agents that can exchange turns equally are
evaluated as having greater fitness.
At first, individuals in each population are initialized
with random weight values.
Then we calculate the fitness of each individual, based on
its performance.

The highest value is given when both agents take their turn
alternately and the agents can predict each other's behaviour.
A one-sided (i.e., role-fixed) behaviour is associated with lower
fitness values. Practically, the fitness of an agent $\bm{a}$ from a population ($A$)
against an agent $\bm{b}$ from the other population ($B$) is calculated as follows.
Below, we define a total fitness $F$ as the sum of two fitnesses associated with
prediction and turn-taking, respectively. When one agent gets behind the other, by definition the other agent
has its turn and the rear scope is specified as $RS$, which is
parameterized by two parameters $r$ and $\phi$ (see Fig. \ref{fig:scope}).
The agent in this state is said to be having its turn and is rewarded.
A spatial position of agent $\bm{b}$ at time step t
is represented by $Pos_{b}(t)$. This is compared with
agent $\bm{a}$'s prediction value $Pos_{a\rightarrow b}$. Therefore the squared difference
(Eq.(11)) is the measure of the precision of agent $\bm{a}$'s prediction.
\begin{eqnarray}
F_{a} &=& \frac{1}{P}\sum^{P}\left(s_{1}\times F_{a}^{turn} +
s_{2}\times F_{a}^{predict}\right),\\
F_{a}^{turn}&=&
\sum^{T}_{t}g_{a}\left( t \right)
\times\sum^{T}_{t}g_{b}\left( t \right),\\
g_{a}( t ) &=& \left\{
\begin{array}{ll}
1 & Pos_{a}(t) \in RS_{b}(t)\\
0 & Pos_{a}(t) \notin RS_{b}(t)\\
\end{array} \right\},\\
F_{a}^{predict}
&=& -\sum^{T}_{t}P_{a}\left(t\right)
\times\sum^{T}_{t}P_{b}\left( t \right),\\
P_{a}( t ) &=& (Pos_{b}(t)-Pos_{a\rightarrow b}(t))^{2}.
\end{eqnarray}
The performance of turn-taking is evaluated for different lengths of time ($T=
500, 1,000$ and $1,500$), so that agents cannot tell when the
evaluation time is over.
Evaluating the turn-taking performance at each GA generation,
we leave the best $E$ individuals in each population and let them reproduce
with specified mutation rates.
The GA proceeds by repeating this procedure, and
the recurrent neural networks evolve. \\
In addition, the following points should be noted.

\subsection{Two Time Scales}

Two time scales exist: the vehicle navigation time scale ($\Delta T_1$), and
the neural computation time scale ($\Delta T_2$). The time
evolution of the vehicle navigation is computed
using the 4th order Runge-Kutta method, where $\Delta T_1$ is set to 0.01.
The basic process is that the neural
net receives the sensor inputs and computes the motor outputs.
By assuming that the vehicle navigation motion is faster
than the internal neural time scale, we take $100 \Delta T_1 = \Delta T_2$.
For simplicity, the neural net produces the outputs every 100 Runge--Kutta
time steps. When the network
structure evolves by GA, the time scale ratio is implicitly reflected in
the net structure. Therefore, we believe that the same behaviour
structure can be obtained, at least qualitatively, for a different scale ratio.

\subsection{Noisy Environment}

Living systems are involved in a fundamentally noisy environment.
We know that our perception has to deal with noisy inputs. However,
it is not possible to discriminate noise from other signals.
We, as living systems, behave adaptively, cooperatively or selfishly
while handling the problem.
Therefore, we simulated the agents' interacting with each other in a noisy environment.
Noises are added to the input neurons at every game step
during each run in the GA.
The strength of noise is provided by uniform random numbers between
zero and almost the maximum distance the agent can move during one game
step.
In the next sections, spatial patterns of turn-taking are studied as simulation results.
If there is no excuse, those patterns are generated under a noise-free environment to
clarify the intrinsic dynamics of the agents.

\section{Simulation Results}

Simulation was performed with a GA using 15 individuals ($P=15, E=4$).
After several thousand GA generations, turn-taking is established
between the two agents. The basic dynamics of the turn-taking was observed as
follows. Two agents adjust their speeds and make turns
automatically to switch from the role of evader to chaser and {\sl vice versa}.
In the following subsections, we investigate the turn-taking
pattern realized from the dynamic repertoire, predictability, adaptability and
evolvability concepts.

\begin{figure}[hptb]
\centerline{\includegraphics[scale=0.3]{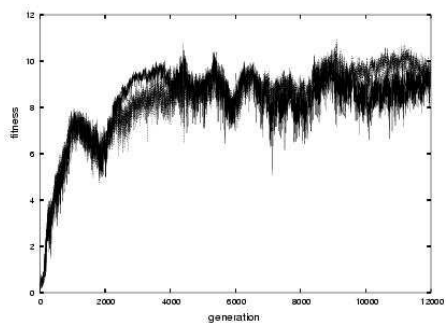}}
 \caption{
Fitness values of the best agents in two populations at each GA
generation for a single run.
}
\label{fig:GA}
\end{figure}

\subsection{Diversity of Dynamic repertoire}

First, the evolutionary algorithm effectively functions to improve the
turn-taking performance. The development of the performance as a function
of GA generations is depicted in Fig. \ref{fig:GA}. The resulting turn-taking patterns are
sensitive to some of the settings. In particular, they are sensitive to the
division of the agent population into two. In previous work,
we encoded the pair of agents' structures on the same gene \cite{ezcaikeg2002}. Then
we encoded them separately but used a single population. That
algorithm can also develop turn-taking behaviour but with much less diversity
than the present algorithm. When the agents are on the same gene, it is difficult
to show diversity as their net structures are too correlated.
With a single population, development of an agent that can take
turns with itself (its relatives) is enhanced. Therefore, there is a strong probability that the dynamics
of the turn-taking may be tuned for self-turn-taking. To avoid
this situation, we used the two-population structure.

\begin{figure}[hptb]
\centerline{\includegraphics[scale=0.7]{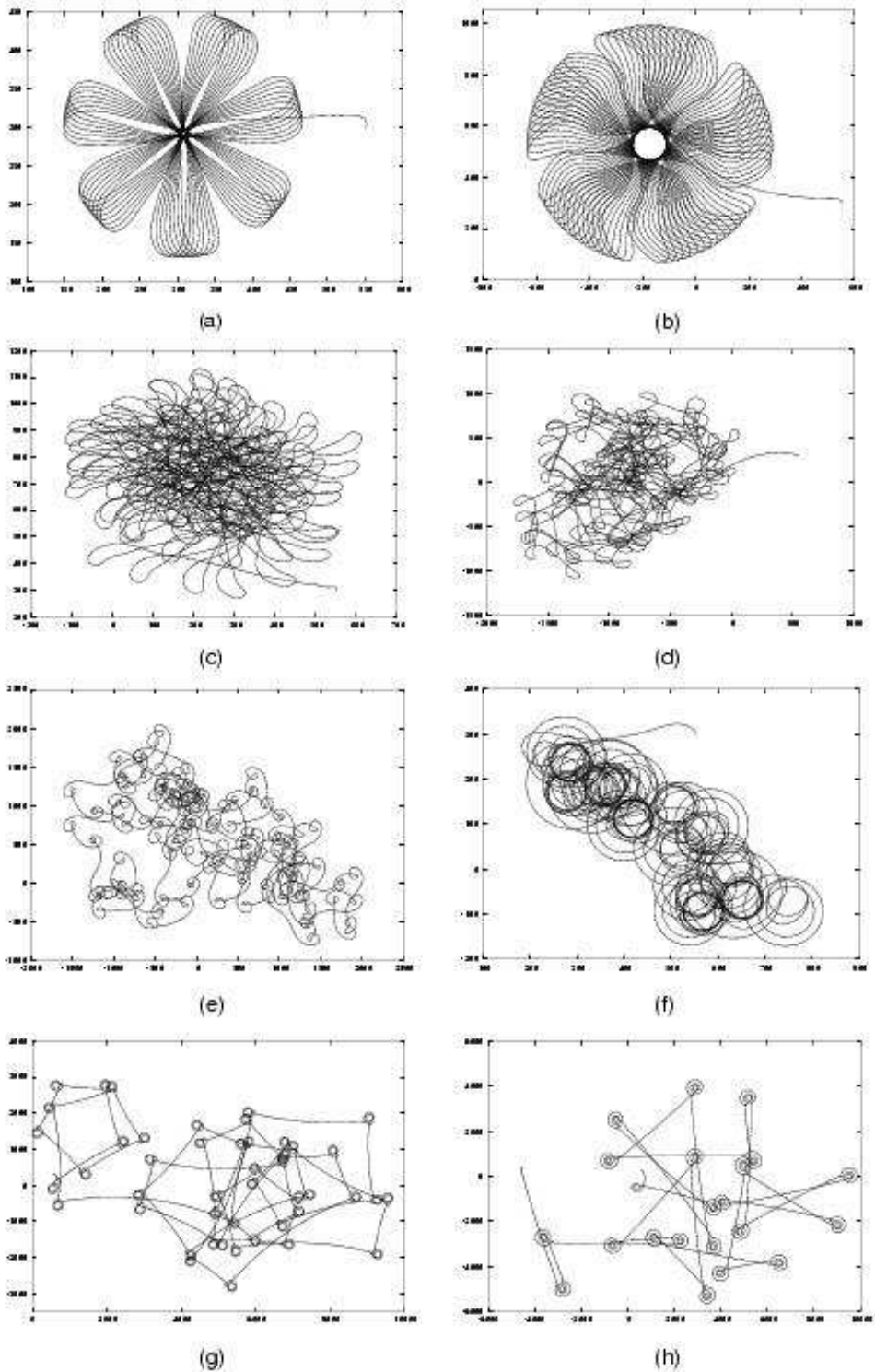}}
\caption{Spatial trails of turn-taking behaviour observed in the simulations.
To clarify the qualitative difference of
turn-taking structures, a spatial trail of only one of the two agents is shown.
The other agent moves around these trails generating similar
trails.
All games in these graphs are started from (550, 300).
(a) and (b) are examples of regular turn-taking behaviour, while the others are examples of
chaotic turn-taking behaviour.
}
\label{fig:orbit}
\end{figure}

Figure \ref{fig:orbit} shows examples of the spatial trails of an agent
from different GA generations with different initial population
structures. For the sake of clarity, a single agent's trail is
depicted. A paired agent tends to show the same trail with different phases.

We can classify these trail patterns approximately into
regular, chaotic and others based on their appearance in space and time.
When spatial trails consist of regular curves, and turns are
exchanged almost periodically (which corresponds to a turning
point on the figures), we call them {\em regular} turn-taking.
On the other hand, if spatial trails have irregular curves with
non-periodic turn-taking, we call them {\em chaotic} turn-taking.
The remaining unclassified patterns are discussed below.

In the earlier GA generations, agents with regular turn-taking
evolve to yield higher performance (Fig. \ref{fig:orbit}(a) and (b)).
The behaviour structure is as follows. One agent follows the other and passes it;
then it slows as does the other agent; then
both agents simultaneously turn around quickly.
This returns the agents to the first phase.
A series of behaviour patterns repeats almost periodically and the
envelope curve of these trails constitutes a circle by fixing the
centre location.
In the later GA generations, more chaotic patterns emerge
(Fig. \ref{fig:orbit} (c) to (h)).
In contrast to the regular patterns, the turns are exchanged in different
places with irregular time intervals. Therefore, the spatio-temporal
pattern becomes chaotic and agents move around the entire space.

\begin{figure}[hptb]
\centerline{\includegraphics[scale=0.5]{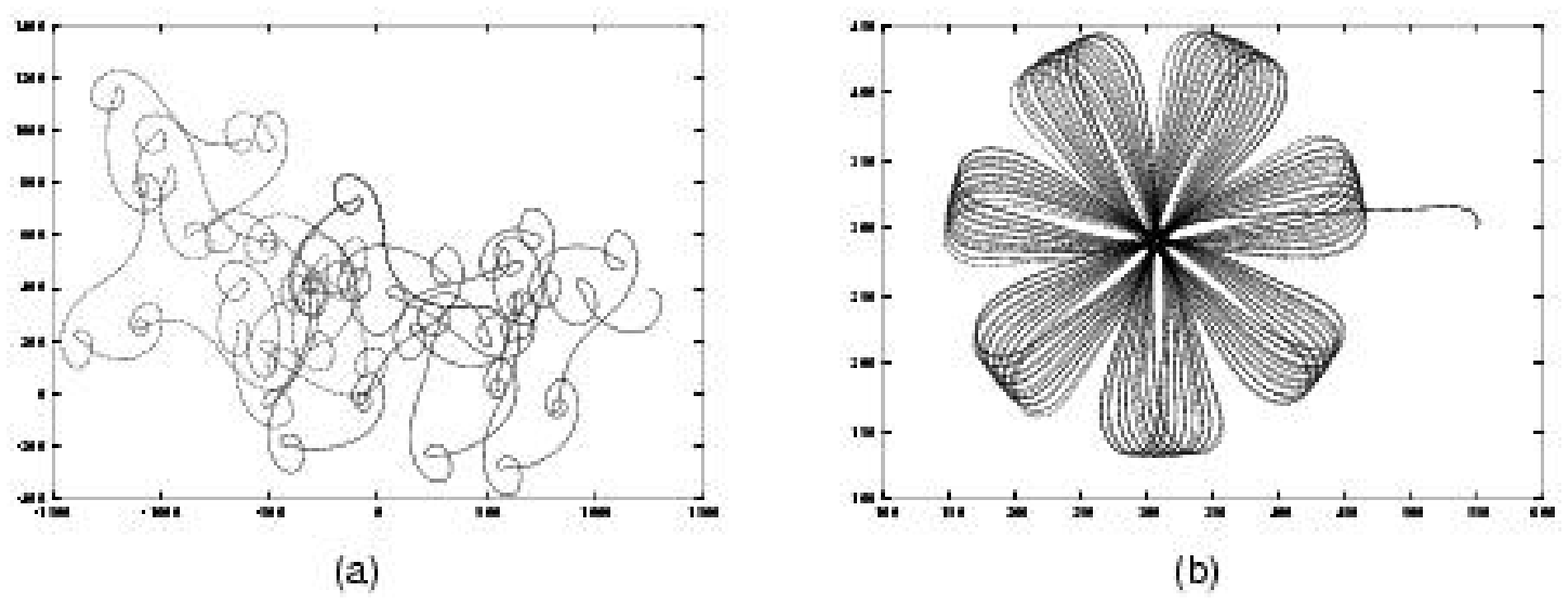}}
\caption{Differences of spatial trails between adaptive agents without noise
(solid) and with noise (dotted) are plotted. They start from the same
initial points, (550, 300).
(a) chaotic turn-taker (b) regular turn-taker
}
\label{fig:difforbit}
\end{figure}

The evolution of turn-taking type from regular to chaotic
is explained as follows. The evolutionary pressure of GA
at first allows the agents to move stably in the noisy environment.
A structured turn-taking behaviour can only be built up on
stable motion dynamics that are insensitive to random noise.
As argued briefly in the introduction, noise and
intentional action is difficult to distinguish when the agents' motions
become chaotic. However, when their actions appear regular, we can interpret that 
the agents can more
easily distinguish noise from the other agent's intentional motion as they 
show different performance with and without partners' adaptive motions (see \S 4.3).
Therefore, the regular type emerges earlier than the chaotic motion.
As shown in Fig. \ref{fig:difforbit}, regular turn-taking occurs at
almost the same spatial location with different noise series. However, the
chaotic type is sensitive to the noise series. The total performance of
turn-taking remains high in both cases.

That is, regular turn-taking pattern suppresses a variety of
dynamic repertoires. By doing so, it becomes robust against sensory
noise. On the other hand, chaotic turn-taking pattern has the potential to develop
dynamic repertoire, and therefore it becomes more adaptive, which is studied
in \S 4.4.

\begin{figure}[hptb]
\centerline{\includegraphics[scale=0.6]{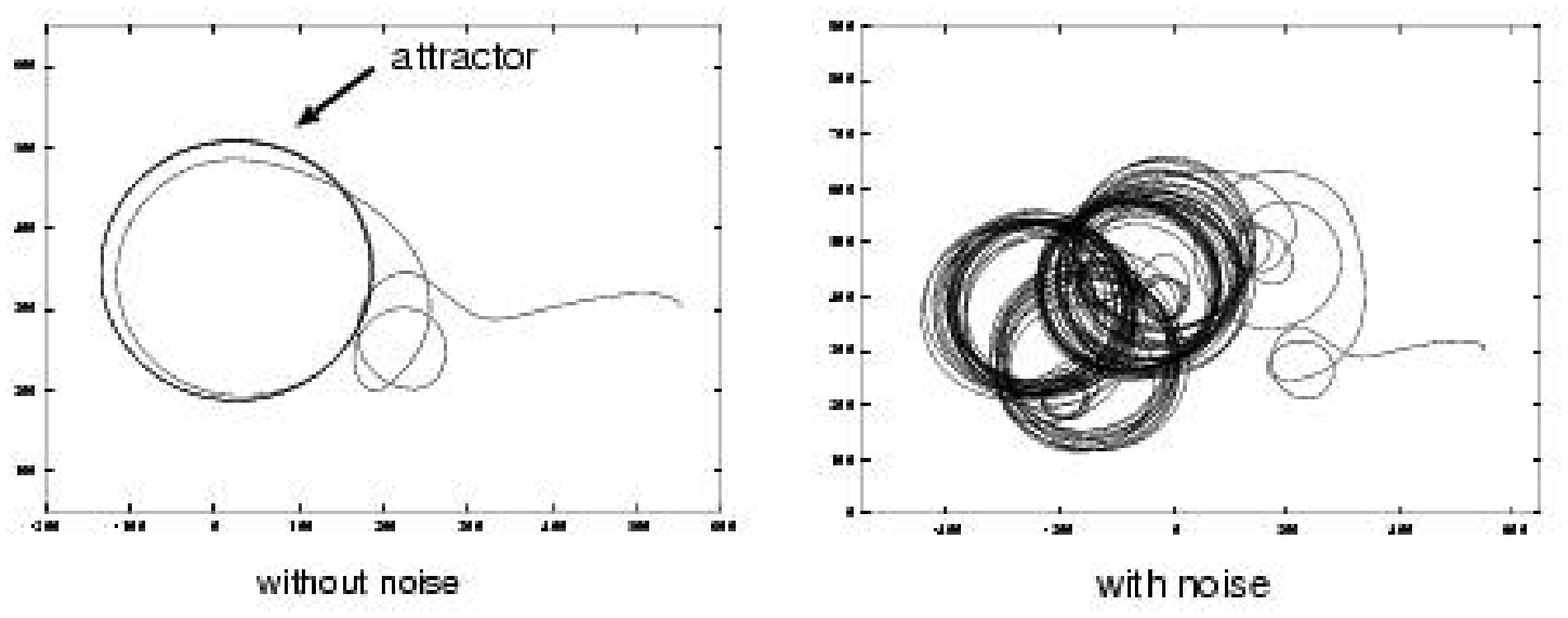}}
\caption{Noise-induced turn-taking behaviour. There is an attractor of
role-fixed behaviour. By adding noise to the agents, an agent can slip out
of the attractor and successfully perform turn-taking.
}
\label{fig:noise-driven}
\end{figure}

\begin{figure}[hptb]
\centerline{\includegraphics[scale=0.3]{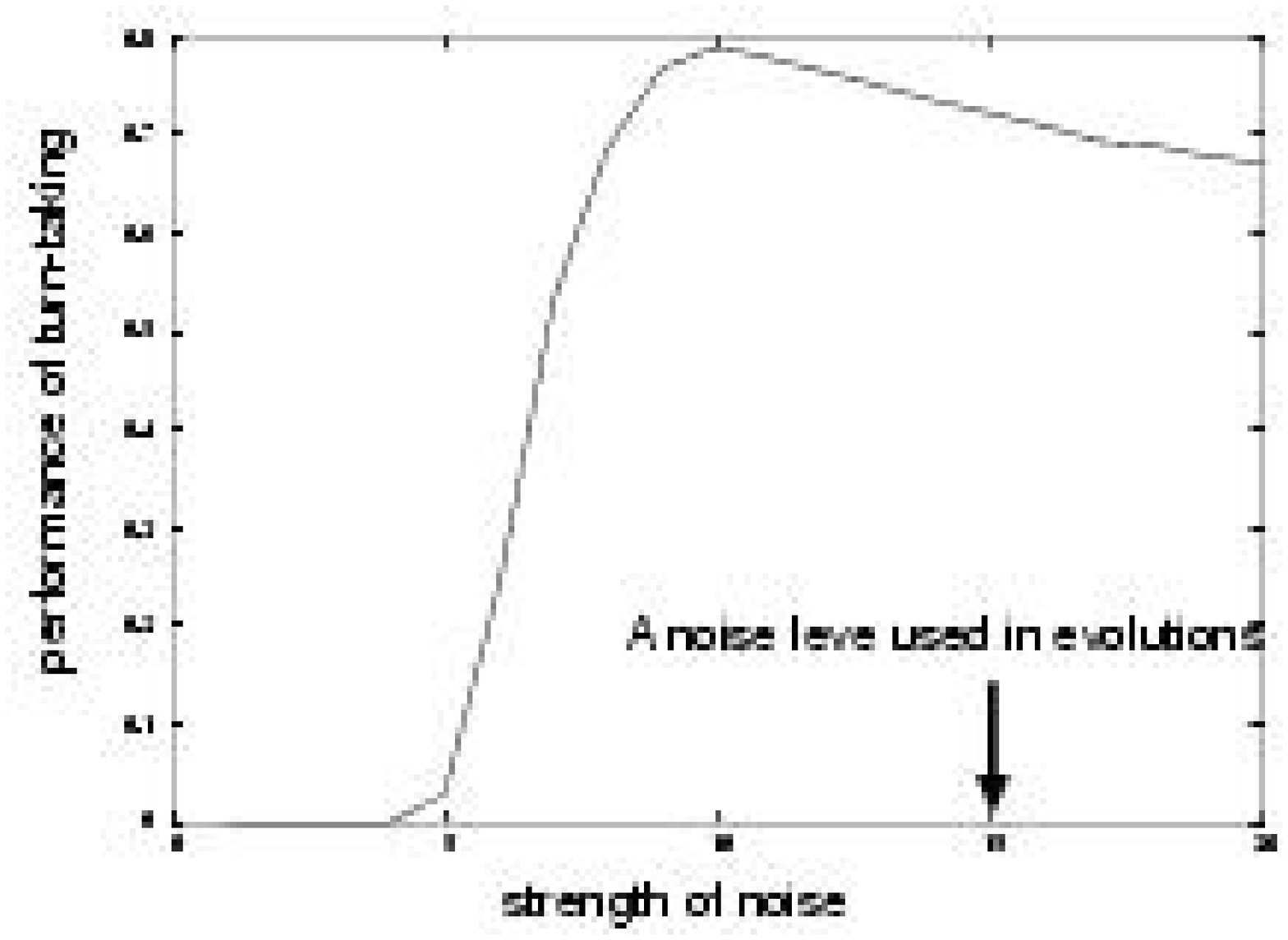}}
\caption{The performance of turn-taking behaviour as a function of noise strength.
Below a certain noise level, agents cannot perform
turn-taking. Above a certain noise level, agents take advantage of
noise to perform turn-taking. This critical noise level is lower than
that used in evolution.
}
\label{fig:noise}
\end{figure}

Intuitively, agents who can take turns in the presence of noise can
take turns perfectly without noise. However, this does not hold for some agents found in
later GA generations.
As shown in Fig. \ref{fig:noise-driven}, agents can only take
turns when there is sensory noise.
We call this phenomenon {\em Noise-induced} turn-taking.
As shown in the figure, there is a strong attractor
to a circular motion without exchanging turns. The two agents have
different neural structures, and the resulting turn-taking behaviour is
generally asymmetrical. Without noise, one agent is never able to take
its turn. In addition, it forms an attractor in the sense that adding a small noise
cannot break this one-sided behaviour. True turn-taking only emerges
above a certain noise level (Fig. \ref{fig:noise}).
In another case, there exist three attractors when there is no sensory noise.
One is that agent A chases the rear of agent B closely.
Another is the opposite, and the last is that in which both agents
chase each other. Each of the three attractors consist of circular orbits.
The transition between attractors is caused by noise.
Without noise, agents are trapped by one of the attractors.

Compared with these noise-induced behaviours, chaotic turn-takers can
spontaneously establish turn-taking behaviour without noise.
Even if noise is introduced into the system, chaotic turn-takers can establish
turn-taking behaviours independent of the low noise level. That is,
they do not utilize noise
but suppress the effect of noise to perform turn-taking. Conversely,
noise-induced turn-takers need noise to perform turn-taking.

\subsection{Prediction Capability and Role Switching}
\begin{figure}[hptb]
\centerline{\includegraphics[scale=0.7]{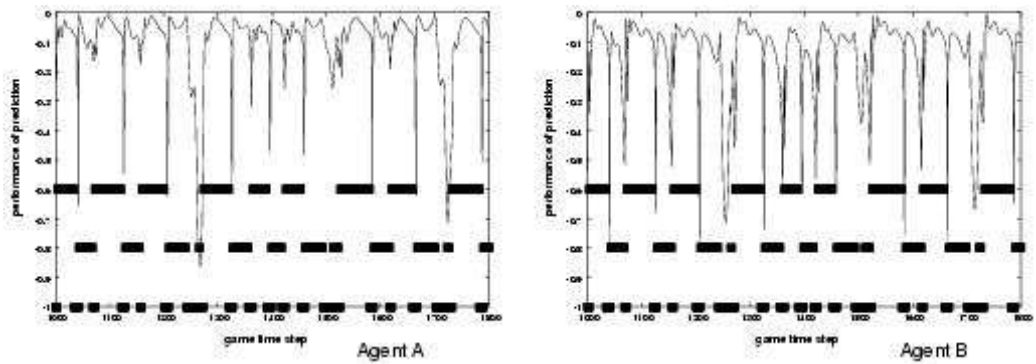}}
\caption{
Prediction (top spiky lines) and turns (line segments)
are drawn for each agent from 10,000 GA generations. A horizontal line
expresses time steps for two agents moving in the two-dimensional arena.
The top two line segments correspond to
turns of the coupled agents. The bottom segments correspond to times when neither agent has a turn.
This shows that the prediction precision decreases sharply when a turn is switched.
}
\label{pred}
\end{figure}

These observations were analysed in terms of prediction capability of agents.
The agents, after thousands of GA generations, are able to predict their
partner's future movements while turn-taking.
Three outputs of the recurrent network simulate the other agent's future
location and heading from the current input.
Fig. \ref{pred}, shows the precision of predictions and the
associated turn-taking patterns.
In earlier GA generations, one agent's prediction is far better than the
other's. In later generations, both predictions are improved. However, through
entire GA generations, the predictions almost
periodically break down when their turns (roles) are exchanged.
As indicated in the figure, the prediction is also perturbed by
noisy inputs. However, the effect is much smaller than that of the other
agent's action.

It should be noted that
these prediction outputs are not designed explicitly to do anything in
generating action sequences.
However, because they depend on the common context neurons that also control
the motion patterns, simulating each other's behaviour and
generating the motor outputs have indirect correlations.
The correlation between prediction breakdown and the turn-taking performance will be
reported elsewhere.

\subsection{Ongoingness of Interactions}

The inherent adaptability of each turn-taking pattern can be studied using
its stability in the presence of noise. In other words, we study an agent's ability to
discriminate between noise and the adaptive behaviour of the other agent.
In this section, we compare the behaviour of ``live interaction'' with ``recorded
interaction''. The ``live interaction'' is normal interaction between
evolved agents, and the ``recorded interaction'' is that between an agent
and a virtual agent, defined below.

First, we selected the two best agents, A and B, from each population.
Turn-taking between these agents was studied without
introduced noise. This is what we term ``live interaction''.
The trails of the agents were recorded during the run.
Then, turn-taking between agent A and the recorded trail of agent
B (i.e., a virtual agent) was
conducted. This is what we term ``recorded interaction''.
We perturb the recorded trail and simulate the changes in
the turn-taking dynamics.

\begin{figure}[hptb]
\centerline{\includegraphics[scale=0.55]{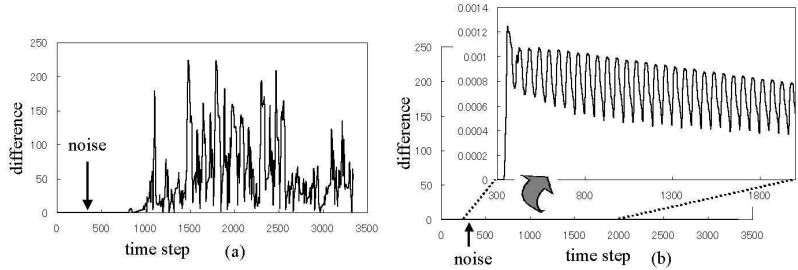}}
\caption{Differences of orbits between agents' trails in a game with
an adaptive agent and a recorded trail.
A small noise is introduced at 340 time steps.
If there is no noise, no difference is observed.
Agents used in (a) and (b) correspond to those in Fig. \ref{fig:orbit}
(e) and (a), respectively.
The difference is amplified if agents fail to establish turn-taking.
}
\label{fig:tracediff}
\end{figure}

Figure \ref{fig:tracediff} (a) shows the growth of a discrepancy between
A-virtual B and A-perturbed virtual B (chaotic turn-takers).
During the initial few hundred steps, no discrepancy was observed.
The behaviours are similar
as shown in the figure. However, a small noise was amplified and
the orbit drastically changed from the original orbit at approximately 800 time steps.
In terms of the turn-taking behaviours, the adaptive agent can no longer
recover harmonization with the perturbed virtual agent.
The agent approaches the trail and tries dynamically
to resume the original turn-taking behaviour.

Another example (the agents at 3,000 generations) is shown in
Fig. \ref{fig:tracediff} (b).
These agents established regular turn-taking.
In this case, the agents could cope with the perturbed virtual agent.
Note that agents that have constructed regular
turn-taking behaviour do not always, but frequently do, have a tendency to cope with a
perturbed virtual agent,
although this varies with the timing and strength of the perturbation.
Sometimes turn-taking behaviour breaks down when more noise is added to the
recorded trail.
However, there are some examples in which turn-taking recovers after a period of discrepancy.

\subsection{Evolution of Adaptability}

\begin{figure}[hptb]
\centerline{\includegraphics[scale=0.4]{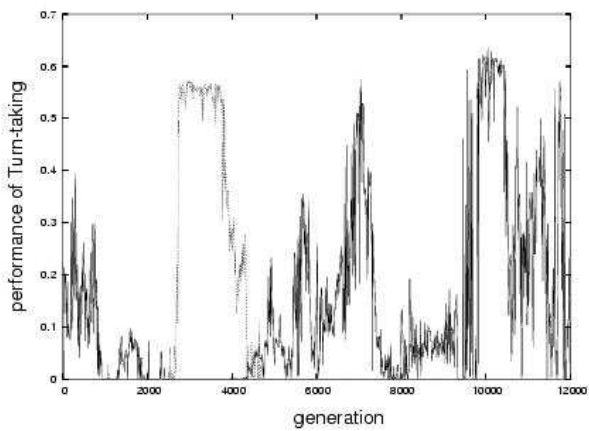}}
\caption{
The best agents from the 10,000 (solid line) and the 3,000 (dashed line) GA
generations are examined with regard to
coupling them with the best agents from different GA generations.
The performance of turn-taking of the newly coupled pair is evaluated for each generation.
Generally the performance is lower than the
original performance of the best pair from each generation, which is
approximately 0.6.
}
\label{10000Vs}
\end{figure}

\begin{figure}[hptb]
\centerline{\includegraphics[scale=0.8]{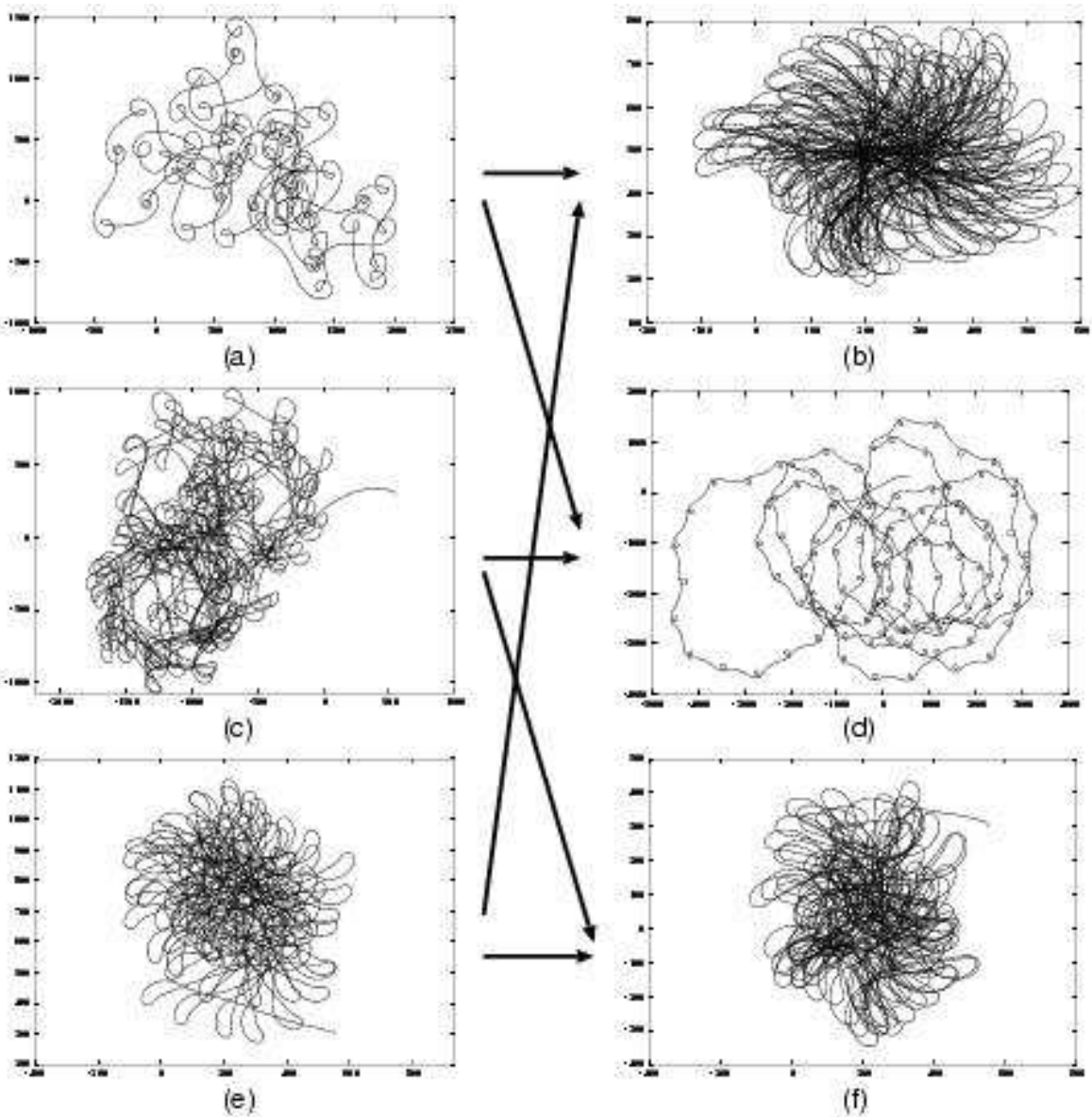}}
\caption{Spatial trails of the original pairs and newly coupled agents at
7,000, 8,000 and 10,000 GA generations. (a), (c) and (e)
show the trails of the original pairs at 10,000, 8,000 and 7,000 generations, respectively.
On the right, newly coupled agents' trails are shown.
(b), (d) and (f) are generated by the best agents at 10,000 vs 7,000, 10,000 vs
8,000, and 8,000 vs 7,000 generations, respectively.
(b) and (f) are similar to the trails generated by the original paired agents from the
7,000 generation.
On the other hand, (d) shows a new trail.
}
\label{orbits2}
\end{figure}

Another novel feature of adaptability was examined.
We show here that adaptability can generate novel dynamics by
constructing new couplings.
We examine the behaviours of new couplings between two agents from
different GA generations as follows.
After the turn-taking performance had attained a satisfactory plateau,
we selected two individuals from different generations
to play. This was to examine how they performed turn-taking
without having the common experience of co-evolution. Taking agents
from generations 10,000 and 3,000 as examples, we evaluated the performances of
the new pairs for each generation (Fig. \ref{10000Vs}). In fact,
the novel pairs often failed to sustain the same performance
as the original pairs. However, the synthesized dynamics often
showed novel structures. The examples can be found in Fig.\ref{orbits2}.
Agents that perform chaotic turn-taking after 10,000, 8,000, and 7,000 generations
(Fig. \ref{orbits2} (a),(c) and (e)) are coupled with agents from each different
generation.
As is seen in the figure, the newly coupled agents also show
chaotic turn-taking but with a different kind of
motion (d).
Coupling of generations 1,000--7,000 and 8,000--7,000 shows a similar pattern to that by the agents from generation
7,000, which is shown in (b) and (f).

In summary, (i) Novel structures sometimes
inherit the original pattern of one of the agents but not always,
(ii) Agents that readily exhibit chaotic turn-taking pattern lose the
original pattern and adapt to the other agent's pattern,
and (iii) conversely, regular turn-takers simply retain their
original pattern and show little adaptability to a new partner.

\begin{figure}[hptb]
\centerline{\includegraphics[scale=0.7]{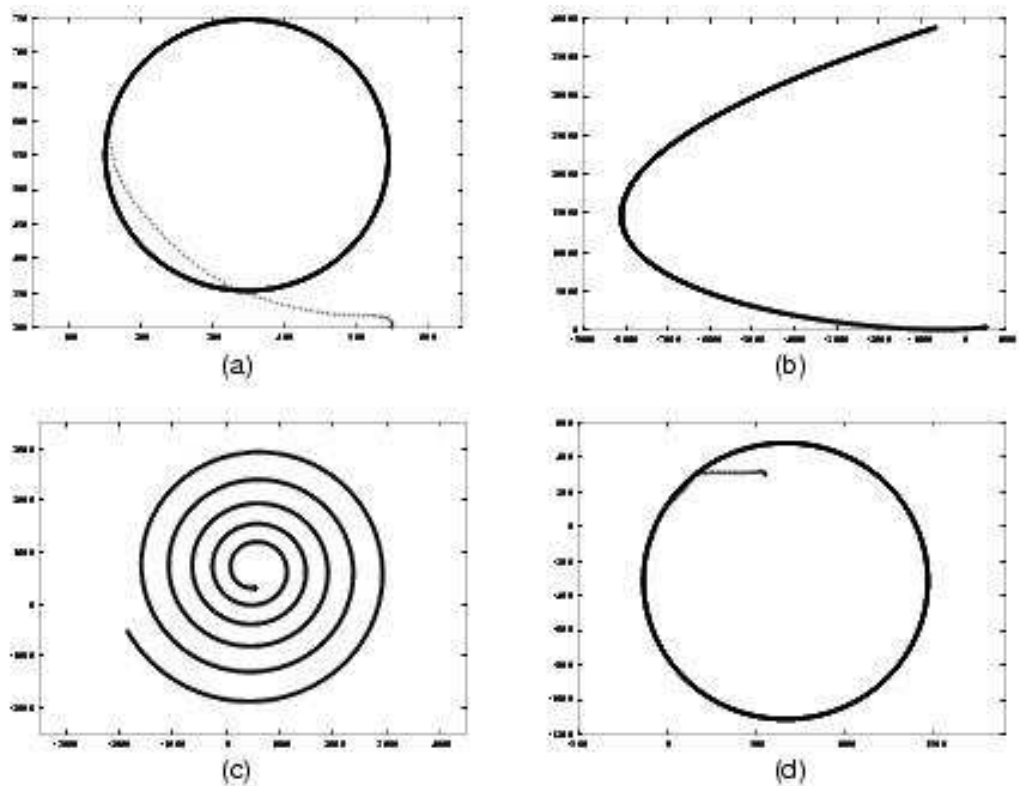}}
\caption{Spatial trails of new couplings of regular and chaotic turn-takers.
(a) 3,000 vs 7,000 (b) 3,000 vs 8,000 (c) 3,000 vs 10,000 (d)
3,000 vs 27,280. One agent always chases the partner, and role changing did not occur.
Convergence of agents' sensors and motors
causes the decrease in behavioural diversity and the interruption of role changing for turn-taking.
}
\label{vsregular}
\end{figure}

\begin{figure}[hptb]
\centerline{\includegraphics[scale=0.8]{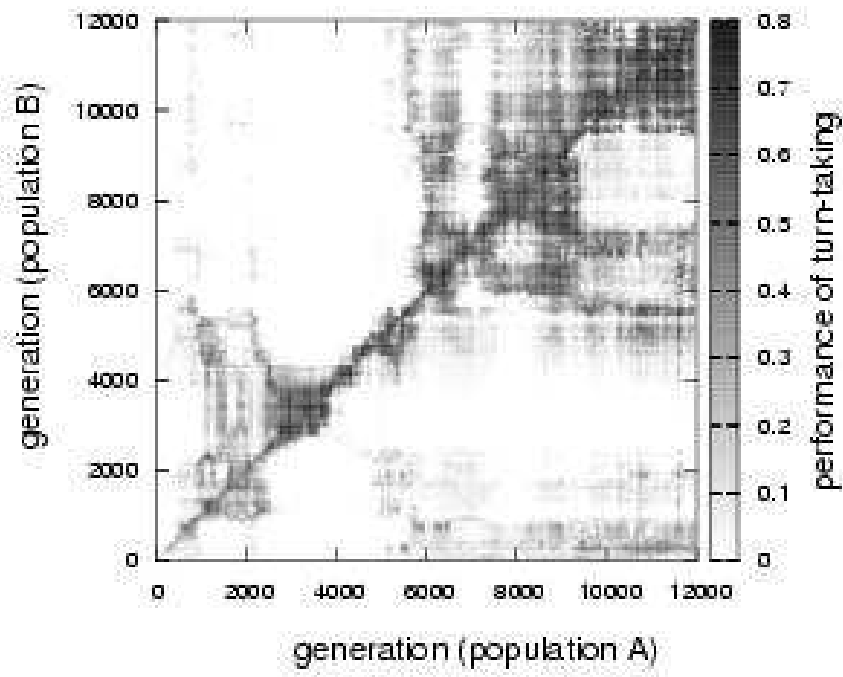}}
\caption{
The performance of turn-taking by new couplings with the best agents from
all generations of two populations. Beyond 6,000 generations, patterns
change from regular to chaotic. The agents after 6,000 generations show
a tendency to be able to perform turn-taking with agents from different generations.
}
\label{All}
\end{figure}

The last point is clearly shown in Fig. \ref{vsregular}.
The regular turn-takers can only achieve higher performance
with agents from near generations (Fig. \ref{10000Vs})
Our hypothesis is that chaotic turn-takers are more adaptive than
regular ones. The observation here confirms the hypothesis, but we should
note that performance sometimes differs significantly between
populations A and B from the same GA generation. Figure \ref{All} illustrates how
turn-taking performance varies from generation to generation.
We deduce from this figure that they are basically symmetrical for
populations A and B. Sometimes there are notable exceptions---e.g., population
A from generation 8,000--10,000 compared with population B from
generation 10,000--12,000. It should also be noted that genetically closer
agents can collaborate better than more distantly related agents. However, qualitatively,
beyond generation 6,000, agents become more adaptive than those of earlier
generations.

The turn-taking pattern resulted from the collaboration of two
agents. Therefore, a neural structure in the body of a single agent
alone cannot explain the turn-taking dynamics.
This is an interesting part of the present study, but at the same time,
a gap between the two agents may develop. That is, when one population becomes
very adaptive against many others, it is not necessary for the other population
to become very adaptive; it may simply become a test data set for
the former population to become ``universal'' turn-takers.
As far as we know, such universal turn-takers are yet to evolve.
Here we notice that chaotic turn-taker is better at eliciting coordinated behaior
from the partner.

It is also worth noting that the ``experience'' of two
agents interacting with each other is a prerequisite for better turn-taking.
The history, or the experience, of how agents have collaborated to perform
turn-taking determines with whom an agent can take turns.
In new pairs, responses of one agent to the other often occur
at the wrong time, whereas the original pairs show complete synchronization of turn-taking.
That is, we insist that it is not the neural structure but
the collaboration of timing and patterning that is responsible for the better
turn-taking behaviour. This is true not only for this special type of
interaction---i.e., turn-taking---but may be true for cognitive interaction in general.
We will argue this point in the final section of this paper.

\section{Discussion}

It was found in the virtual agents experiment (\S 4.3), that
chaotic turn-takers are much more sensitive to
the difference between live and recorded inputs.
Their turn-taking patterns are driven by the ongoing interaction.
On the other hand, regular turn-takers are relatively insensitive to the difference.
As evolution continues, chaotic turn-taking replaces
regular turn-taking in the GA simulations.
This may be due to regular turn-takers' being less adaptive
than chaotic turn-takers in the sense that they can only cope with
fewer agents. This is clearly seen in the
new coupling experiment \S 4.4. The coupling with regular turn-takers only
generates circular patterns but chaotic turn-takers show various
patterns.
In summary, we claim that
chaotic turn-taking is less robust in the presence of noise but has more adaptability,
compared with regular turn-taking.

This complementary relationship between adaptability and robustness has
some implications in some empirical experiments. Let us introduce
Trevarthen's double-monitor experiments
between a baby-infant and its mother \cite{tre77,tre}, and Nadel's mutual
imitation experiments \cite{nadel}. In Trevarthen's experiment,
mother and baby-infant only communicate through videos that display
their faces to each other. For the baby-infant to engage with the mother, correct
style and timing are required.
If the recorded video of the mother is displayed to the baby-infant,
the baby-infant becomes withdrawn and depressed.
Nadel studied how the mutual imitation game
progresses between children and discussed a non-affordant means of using objects
to trigger the interaction. Children regularly switch between
the roles of imitating and being imitated, by having new
imitation patterns.

Trevarthen's experiments show that it is not necessarily
important for the baby-infant that the mother be displayed on the monitor.
It can be assumed that the most important clue during interactions
is the ongoing anticipation of a partner. The baby-infant performs some actions and
anticipates the mother's reactions reflecting the baby-infant's actions, and this
is also true with respect to the mother's anticipation of the baby-infant.
Interactions in social behaviour, including turn-taking, can be
established when these anticipations are mutually formed dynamically.
Furthermore, it is shown by Nadel's experiment that an affordant way of using
objects can maintain interaction---i.e. some form of
novelty/unpredictability is required.
In our simulations, when an agent calculates outputs, this calculation
simultaneously affects the internal dynamics.
That is, the actions performed form its internal dynamics as much as actions
form anticipations in the statement above. The agent receives
a partner's actions as inputs that reflect the agent's own actions.
We maintain that turn-taking is established when these structures are
mutually organized. Turn-taking is therefore broken in the simulation with virtual
agents. However, our simulations also show that unpredictability is found
when turn-taking occurs.
We therefore claim that mutually adaptive coupling of actions and internal
dynamics between agents is essential for the establishment of
cognitive interaction, which may be related to intersubjectivity.

{\bf Acknowledgements:} \hspace{0.1cm}
We thank Ryoko Uno and Gentaro Morimoto for their fruitful discussions.
This work is partially supported by
Grant-in aid (No. 09640454 and No. 13-10950)
and also by a grant-in-aid from The 21st Century COE
(Center of Excellence) program (Research Center for Integrated Science) of the Ministry of
Education, Culture, Sports, Science, and Technology, Japan.


\begin{thebibliography}{99}
\bibitem{beer1997}
Beer, R.D. (1997).
The dynamics of adaptive behavior: A research program.
Robotics and Autonomous Systems 20, 257--289
%
\bibitem{vehicles}
Braitenberg, V. (1984).
Vehicles: Experiments in Synthetic Psychology.
Cambridge, MA. MIT Press
%
\bibitem{cliff}
Cliff, D., Miller, G.F. (1996).
Co-evolution of Pursuit and Evasion II: Simulation Methods and Results.
From Animals to Animats 4.
Cambridge, MA. MIT Press, 506--515
%
\bibitem{kerstin1995}
Dautenhahn, K. (1995).
Getting to know each other - artificial social intelligence for
autonomous robots, Robotics and Autonomous Systems 16, 333-356
%
\bibitem{kerstin}
Dautenhahn, K. (1999).
Embodiment and Interaction in Socially Intelligent Life-Like Agents.
Springer Lecture Notes in Artificial Intelligence Volume 1562 Springer, 102--142
%
\bibitem{paolo}
Di Paolo, E.A. (2000).
Behavioral coordination, structural congruence and entrainment in a
simulation of acoustically coupled agents.
Adaptive Behavior 8:1, 25--46
%
\bibitem{ezcaikeg2002}
Iizuka, H., Ikegami, T. (2002).
Simulating Turn-taking Behaviours with Coupled Dynamical Recognizers.
The Proceedings of Artificial Life 8, Standish et al.(eds),
Cambridge, MA. MIT Press, 319--328
%
\bibitem{timt98}
Ikegami, T., Taiji, M. (1998).
Structures of Possible Worlds in a Game of Players with Internal Models.
Acta Polytechnica Scandinavica Ma. 91, 283--292
%
\bibitem{timt99}
Ikegami, T., Taiji, M. (1999).
Imitation and Cooperation in Coupled Dynamical Recognizers.
Advances in Artificial Life, Springer-Verlag, 545--554
%
\bibitem{ikeggenta}
Ikegami, T., Morimoto, G. (2003).
Chaotic Itinerancy in Coupled Dynamical Recognizers.
Chaos: An Interdisciplinary Journal of Nonlinear Science,
Volume 13, Issue 3, 1133-1147
%
\bibitem{ikegezca}
Ikegami, T., Iizuka, H. (2003).
Joint Attention and Dynamics Repertoire in Coupled Dynamical Recognizers.
The Proceedings of Imitation in Animals and Artifacts II, 125--130
%
\bibitem{acper}
Marocco, D., Floreano, D. (2002).
Active Vision and Feature Selection in Evolutionary Behavioral Systems.
In Hallam, J., Floreano, D. Hayes, G. and Meyer, J. (Eds)
From Animals to Animats 7.
Cambridge, MA. MIT Press, 247--255
%
\bibitem{nadel}	
Nadel J., Revel A. (2003).
How to Build an Imitator?
The Proceedings of Imitation in Animals and Artifacts II, 120--124
%
\bibitem{evorobo}
Nolfi S., Floreano D. (2000).
Evolutionary Robotics: The Biology, Intelligence, and Technology of
Self-Organizing Machines.
Cambridge, MA. MIT press
%
\bibitem{understanding}
Pfeifer, R., Scheier, C. (1999).
Understanding intelligence.
Cambridge, MA. MIT-press
%
\bibitem{pollack}
Pollack, J.B. (1991).
The induction of dynamical recognizers.
Machine Learning, Vol. 7, 227--252
%
\bibitem{reynolds}
Reynolds, C.W. (1995).
Competition, Co-evolution and the Game of Tag.
Artificial Life IV, Brooks \& Maes (eds),
Cambridge, MA. MIT Press, 59--69
%
\bibitem{scassellati}
Scassellati, B. (1999).
Imitation and Mechanisms of Joint Attention: A Developmental Structure
for Building Social Skills on a Humanoid Robot. Computation for
Metaphors, Analogy and Agents, Vol. 1562 of Springer Lecture
Notes in Artificial Intelligence, Springer-Verlag
%
\bibitem{classification}
Scheier C., Pfeifer R. (1995).
Classification as sensory-motor coordination.
Proceeding of 3rd European Conference on Artificial Life, 656--667
%
\bibitem{tani}
Tani J. (1996).
Model-based Learning for Mobile Robot Navigation from the Dynamical
Systems Perspective.
IEEE Trans. on System, Man and Cybernetics Part B (Special Issue on
Robot Learning), Vol. 26, No.3, 421--436
\bibitem{tre77}
Trevarthen, C. (1977).
Descriptive Analyses of Infant Communicative
Behaviour. In: Studies in Mother--Infant Interaction,
H.R. Schaffer (ed.), London: Academic Press
%
\bibitem{tre}
Trevarthen, C. (1993).
The Self Born in Intersubjectivity: The Psychology of an Infant
Communicating.
The Perceived Self, U. Neisser(ed.), Cambridge University Press,
121--173
%
\bibitem{walter}
Walter W.G. (1950).
An Imitation of Life.
Scientific American, 182(5), 42--45
%
\bibitem{walter51}
Walter W.G. (1951).
A Machine that Learns.
Scientific American, 185(2), 60--63
%
%
\end{thebibliography}
\end{document}